\documentclass[12pt]{article}
\usepackage{graphicx}
\usepackage{chicago}
\usepackage{geometry} 
\providecommand{\keywords}[1]
{
  \small
  \textbf{\textit{Keywords---}} #1
}

\title{A Deep Learning Model for Forecasting Global Monthly Mean Sea Surface Temperature Anomalies} 
\author{John Taylor\,$^{1,2,3,*}$ and Ming Feng\,$^{4,5}$}

\begin{document}

\maketitle

\noindent
\textsuperscript{1}CSIRO Data61, Canberra, Australian Capital Territory, Australia\\
\textsuperscript{2}College of Engineering and Computer Science, The Australian National University, Canberra, Australia\\
\textsuperscript{3}Defence Science and Technology Group, Department of Defence, Canberra Australia\\
\textsuperscript{4}Oceans and Atmosphere, CSIRO, Perth, WA, Australia \\
\textsuperscript{5}Centre for Southern Hemisphere Oceans Research, CSIRO, Hobart, TAS, Australia\\ \\

\textsuperscript{*}Corresponding Author Address: John Ashley Taylor, CSIRO Data61, CSIRO Synergy Building, Clunies Ross Street, Black Mountain, Canberra ACT 2601.\\
E-mail: John.Taylor@data61.csiro.au

\pagebreak

\section*{Abstract}


Sea surface temperature (SST) variability plays a key role in the global weather and climate system, with phenomena such as El Ni\~{n}o-Southern Oscillation regarded as a major source of interannual climate variability at the global scale. The ability to be able to make long-range forecasts of sea surface temperature anomalies, especially those associated with extreme marine heatwave events, has potentially significant economic and societal benefits. We have developed a deep learning time series prediction model (Unet-LSTM) based on more than 70 years (1950-2021) of ECMWF ERA5 monthly mean sea surface temperature and 2-metre air temperature data. The Unet-LSTM model is able to learn the underlying physics driving the temporal evolution of the 2-dimensional global sea surface temperatures. The model accurately predicts sea surface temperatures over a 24 month period with a root mean square error remaining below 0.75$^\circ$C for all predicted months. We have also investigated the ability of the model to predict sea surface temperature anomalies in the Ni\~{n}o3.4 region, as well as a number of marine heatwave hot spots over the past decade. Model predictions of the Ni\~{n}o3.4 index allow us to capture the strong 2010-11 La Ni\~{n}a, 2009-10 El Nino and the 2015-16 extreme El  Ni\~{n}o up to 24 months in advance. It also shows long lead prediction skills for the northeast Pacific marine heatwave, the Blob. However, the prediction of the marine heatwaves in the southeast Indian Ocean, the Ningaloo Ni\~{n}o, shows limited skill. These results indicate the significant potential of data driven methods to yield long-range predictions of sea surface temperature anomalies.

\keywords{sea surface temperatures, SST, Ni\~{n}o3.4, deep learning, machine learning, AI, Unet-LSTM, ERA5} 

\section*{Introduction}

Modes of inter-annual climate variability, such as El Ni\~{n}o-Southern Oscillation (ENSO) and Indian Ocean Dipole (IOD), are known to modulate the global sea surface temperature (SST) variability and the marine heatwave frequency, duration, and intensity~\cite{McPhaden2006},~\cite{Saji2003},~\cite{Holbrook2019}. The climate modes influence SST variability locally and remotely, mostly through atmospheric teleconnection. Ocean circulation and large-scale oceanic waves also transmit climate signals to remote regions. SST variability and marine heatwave characteristics are also influenced by regional atmospheric and oceanic dynamics and coupled processes. The Blob marine heatwaves in the northeast Pacific during 2013-2015 and in 2019 have been attributed to anomalous atmospheric pressure systems ~\cite{Bond2015}~\cite{Amaya2020}. The teleconnection between the equatorial and northeast Pacific may be one of the key drivers to sustain the Blob warming over a multi-year period ~\cite{DiLorenzo2016}. Marine heatwaves off the west coast of Australia, the Ningaloo Ni\~{n}o, are due to both oceanic and atmospheric teleconnection from the equatorial Pacific ~\cite{Feng2013} and the local air-sea coupling ~\cite{Kataoka2014}, ~\cite{Tozuka2021}. 

A timely forecast of global SST anomalies helps marine and terrestrial resource managers to mitigate potential risks from extreme climatic events. The prediction of ENSO and IOD events, indexed with equatorial SST anomalies in the equatorial regions, is important to forecasting rainfall, drought, and bushfire variability around the globe. Coupled ocean-atmosphere models have been used to forecast global SST variability. Most of the skill assessment has been for the inter-annual climate modes, with 3-4 season forecasting skills for ENSO and 1-2 seasons for IOD~\cite{Stockdale2011}.  Regional SST variability forecast has limited skills and is highly regionally dependent (e.g. ~\cite{Spillman2021}), which is to some extent due to coupled ocean-atmosphere models not properly capturing important regional coupled processes driving the SST anomalies (e.g. ~\cite{Doi2013}).

Coupled model outputs have also been used to train machine learning models to assess the predictability of the climate modes. A convolutional neural network (CNN) model has been proved to have a long-lead prediction skill (up to 18-month) for December-February Ni\~{n}o3.4 SST - an index for ENSO variability, trained by SST and upper ocean heat content anomalies from coupled models ~\cite{Ham2019}. Similarly, an artificial neural network model has been trained to forecast the SST variations at the peak season of the IOD events ~\cite{Ratnam2020}. Given the phase-locking characteristics of the climate modes, these models aim to make single-season predictions, and for a single climate index. SST variability outside the ENSO and IOD regions also show some seasonal phase-locking, such as the Ningaloo Ni\~{n}o marine heatwaves are phase-locked to austral summer ~\cite{Kataoka2014},~\cite{Feng2015}. A CNN architecture trained with climate model outputs has been developed to forecast SST and marine heatwave variability around Australia (Boschetti et al. 2022, personal communication). 

Complex spatio-temporal variations of the climate modes have been recognised, such as the complexity in ENSO dynamics and predictability ~\cite{Timmermann2018}. The SST warming during the 2009-2010 El Ni\~{n}o, the peak SST warming occurred in the central Pacific, so that the event is being classified as a central Pacific El Ni\~{n}o, as compared with the more traditional 2015-16 El Ni\~{n}o when the extreme warming was more located in the eastern equatorial Pacific. There is a variety of temporal variability in the climate modes. Marine heatwaves across the global ocean have diverse spatial variability and are to some extent not tightly phase-locked with seasons~\cite{Gupta2020}. Such as the Blob marine heatwave can occur in different seasons ~\cite{Amaya2020}, whereas the Ningaloo  Ni\~{n}o has substantial spatial variations among different events ~\cite{Feng2015}. A all-season model has been proposed, arguing that it would overcome some arbitrary fluctuations in the predictions at different lead times. Still, the prediction aims for a single index, the Ni\~{n}o3.4~\cite{Ham2021}. Thus, it is important to explore machine learning models which can predict the full seasonal cycle and the spatial patterns of SST anomalies. 

In this study, we propose a new deep learning modeling framework to forecast monthly global SST, using a Unet-LSTM convolutional encoder-decoder neural network ~\cite{Taylor2021}, which has been proven to have better prediction skills while using fewer parameters, compared with other deep learning architectures ~\cite{Larraondo2019}. We train the model with observed (reanalysis) SST and surface air temperature data over the past 7 decades, to demonstrate potential long lead predictions for SST variability in the tropical-subtropical oceans. We present the methodology and examine the predictability of SST anomalies in the equatorial Pacific and the Blob region in detail, whereas full exploration of the machine learning model and SST predictability will be presented in a separate work.

\section*{Materials and Methods}

\subsection*{Dataset}

The ERA5 data set~\cite{Hersbach2020} provides monthly estimates, currently commencing in 1979, of many atmospheric, land and oceanic variables at global scale with a spatial resolution of $0.25 ^\circ$, $\approx 30$km. An ERA5  preliminary analysis commencing in 1950 and covering the period up to December 1978 is also available. The ERA5 data set includes surface variables, including SST, and atmospheric variables computed on 137 levels to a height of ~80 km.  ERA5 dataset was created by combining a comprehensive set of historical meteorological observations with a sophisticated data assimilation and modelling workflow developed by ECMWF. We use a replica of the ERA5 data set available at the National Computational Infrastructure (NCI)~\cite{nci2020}. ERA5 data can also be obtained on request from ECMWF’s meteorological data archive and retrieval system (MARS).
 
\begin{figure}[h!]
\begin{center}
\includegraphics[width=17.5cm]{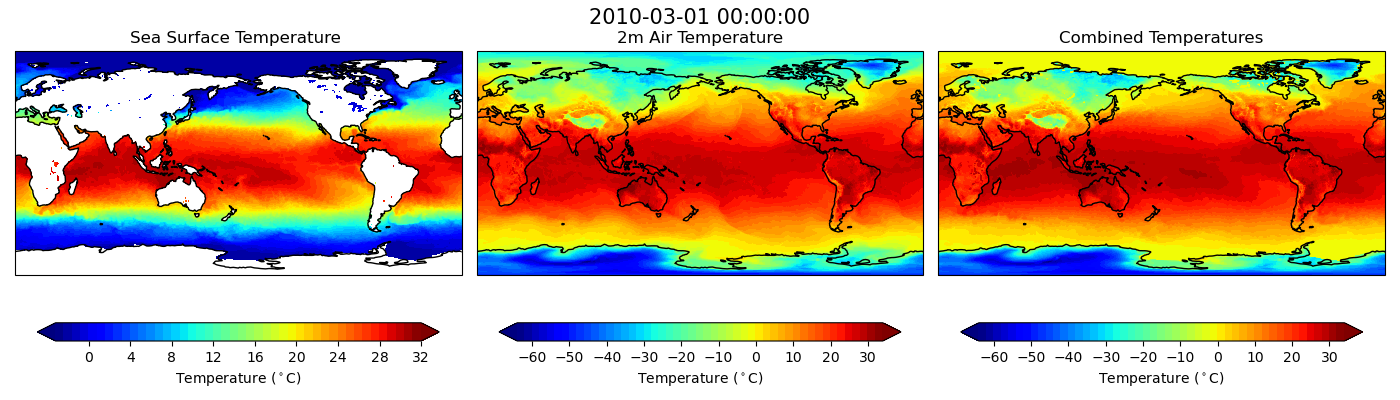}
\end{center}
\caption{An illustration of the ERA5 monthly mean SST, 2-metre air temperatures and the combined temperatures where the 2-metre air temperature is used only over the land surface for March 2010. The correlation coefficient between the SST and 2-metre air temperatures, computed over the ocean only, is 0.92 in this example.}\label{fig:1}
\end{figure}

Our experiments use SST (ERA5 label 'sst') and 2-metre atmospheric temperature (ERA5 label 't2m') variables drawn from the ERA5 data set. As both the SST and 2-metre atmospheric temperature data are from the same ERA5 reanalysis data set they are physically consistent within the limits of the reanalysis system. Figure~\ref{fig:1} provides an example of the ERA5 monthly mean SST, 2-metre air temperatures and the combined temperatures for March 2010. We start with the full global data set with latitude and longitude dimensions of [720,1440]. The temporal domain data span January 1950 until May 2021, with a temporal resolution of 1 month, a total of 857 months. 

The convolutional layers used in our model require complete grids of data, ideally in dimensions that are powers of 2 to avoid the need for padding at the boundaries. To satisfy this requirement we combine the SST data over the ocean grid points with 2 metre atmospheric temperature over the land surface grid points to yield a global grid without masked region over the land surface. Using the Climate data Operators (CDO) software package~\cite{Schulzweida2019a}, we first average the [720,1440] data set to a 1x$1 ^\circ$ grid [360,720] and then use a bilinear interpolation to a [64,128] latitude (-$64 ^\circ$S to $62 ^\circ$N in $2 ^\circ$ increments) and longitude (-$180 ^\circ$S to $180 ^\circ$N in $2.8125 ^\circ$ increments) grid. Finally we normalise the data, as we found that using the normalised data significantly improved the model training performance.The resulting surface temperature data are represented as a three-dimensional numerical array with shape [857, 64, 128] corresponding to dimensions [time, latitude, longitude]. Input data used for training the model were selected as a moving window using 12 time steps (1 year), which captures the seasonal cycle in SST, as this was found to yield the best model predictions of SST.

\subsection*{Models}

We apply a similar deep learning modelling architecture, referred to as Unet-LSTM~\cite{Taylor2021a},  as applied in previous modelling studies~\cite{Taylor2021} except we do not include the batch normalisation layers after each convolution layer as adding this layer did not improve the model fit. We also modify the hyperparameters in order to obtain the model with the best fit.  The Unet-LSTM convolutional encoder-decoder neural network delivers pixel-wise semantic segmentation that enables us to generate quantitative estimates of meteorological variables of interest such as SST at each latitude-longitude grid-point [64,128]. In order to make forecasts of 2-D fields the Unet-LSTM includes a 2-D convolutional long short-term memory (LSTM) layer~\cite{Hochreiter1997}. Other examples of the application of  convolutional encoder-decoder neural networks approaches include SegNet~\cite{Badrinarayanan2017}, VGG16~\cite{Simonyan2015}, and U-net~\cite{Ronneberger2015}. Previous work by~\cite{Larraondo2019} investigated the application of SegNet, VGG16 and U-net to the prediction of precipitation fields and concluded that U-net delivered the best estimates of precipitation while employing significantly fewer model parameters. Based on these advantages and our successful application of our U-net based model in previous studies~\cite{Taylor2021}to the prediction of surface precipitation and forecasting 500 hPa geopotenial height, we have adopted the Unet-LSTM model~\cite{Taylor2021}, as the underlying model architecture for our study. The Unet-LSTM model code is available here~\cite{Taylor2021a}.

\subsection*{Methodology}

\begin{figure}[h!]
\begin{center}
\includegraphics[width=12.5cm, angle=-90]{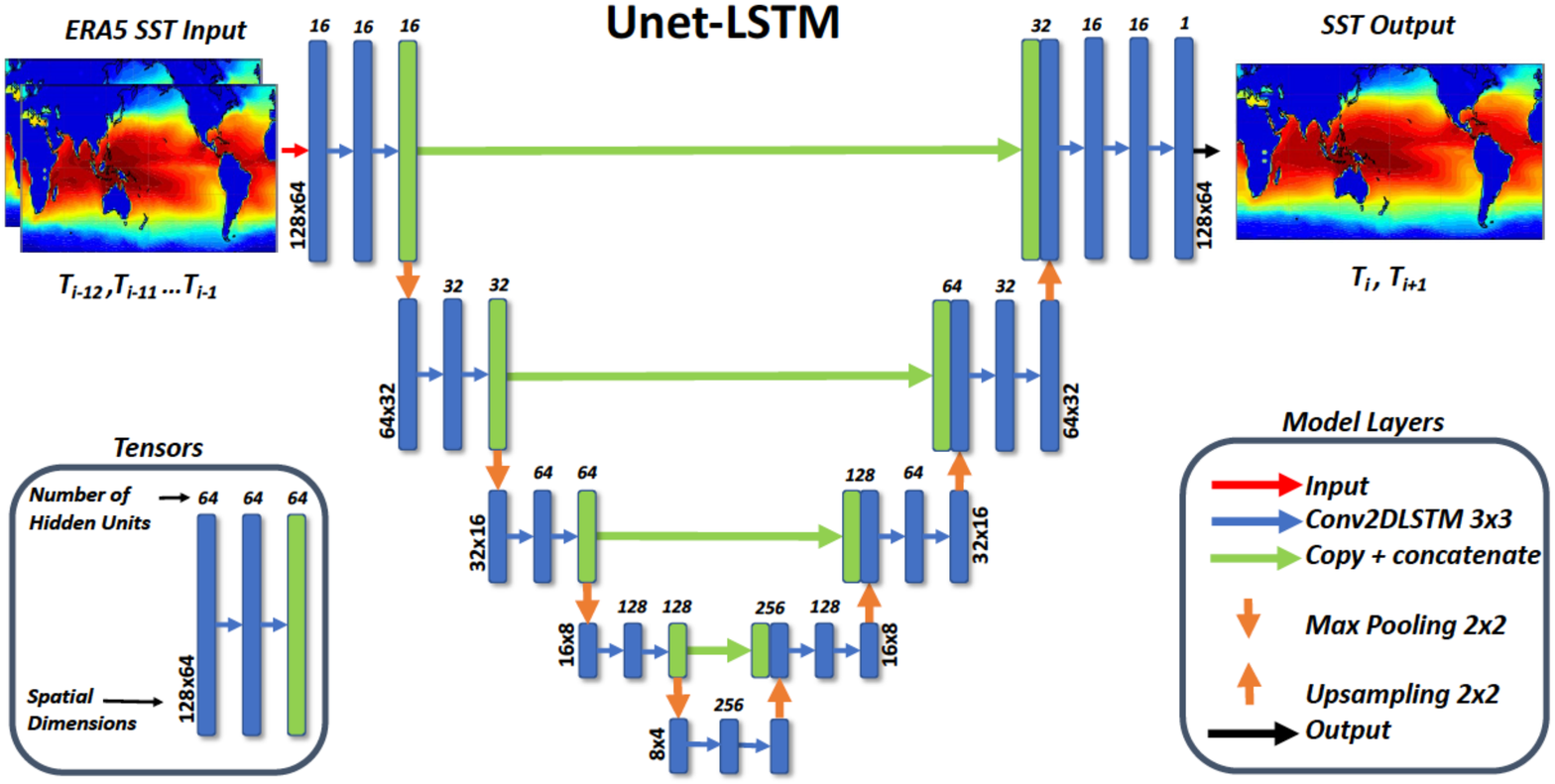}
\end{center}
\caption{A summary of the architecture of the CNN model, referred to as the Unet-LSTM, that we have trained to forecast 2-D SST fields. The model shown was implemented in Python using the TensorFlow and Keras 2.4 API.}\label{fig:2}
\end{figure}

The CNN model described in Figure~\ref{fig:2} was written in Python using the TensorFlow~\cite{abadi2015} and Keras APIs ~\cite{Chollet2015}.  We use Horovod~\cite{Sergeev2018} to implement a data-parallel model. We selected the Adam optimiser ~\cite{Kingma2017} with a learning rate of 0.003 and a learning rate warmup over the first 5 epochs. The higher learning rate and the warmup improved model fitting on the larger batch sizes when using multiple GPUs. We chose a batch size of 4 that yields the best model fit for forecasting SST. The total batch size when using Horovod on multiple GPUs is the number of GPUs multiplied by the batch size on each GPU. The total batch size is therefore a function of the number of GPUs used in model training. For this problem we use 4 Nvidia V100 GPUs each with 32 GB of memory making up one node on the NCI Gadi computer.

\begin{figure}[h!]
\begin{center}
\includegraphics[width=17.5cm]{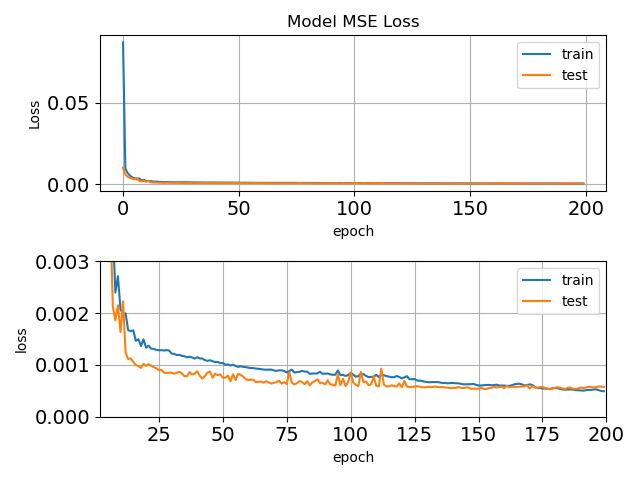}
\end{center}
\caption{The MSE error calculated from a comparison of the model predictions with the training (train) and the validation (test) data sets over a 200 epoch training period.  We see a rapid convergence of the model on the minimum MSE value. The upper panel shows the MSE values over the full 200 epochs. In order to show greater detail of the model fitting process, the lower panel shows the same MSE values at higher resolution and focused on the tail of the model training. Note the different y-axis scales between the two panels.}\label{fig:3}
\end{figure}

We use 760 of the available time steps from January 1950 to April 2013 for training and the remaining 97 time steps for validation and testing. The CNN model uses the prior 12 months of SST data in order to predict the following 2 months of SST data. Note that selecting a longer prediction time period results in errors accumulating over a longer prediction window. We used the tanh activation function, set the kernel, bias and recurrent L2 regularisation value at 10\textsuperscript{-8} and run the model training for 200 epochs saving only the model with the minimum MSE value, then output the final model and report the resulting MSE values. We found that 200 epochs ensured that the MSE value always reached a minimum without overfitting. Figure~\ref{fig:3} shows the MSE value calculated from the model fit to the training and test data sets over 200 epochs. Using the saved model we then make model predictions (inference) using an autoregressive approach for up to 24 months.

In order to efficiently load the model training data using a data-parallel approach we distribute the model data required by each GPU onto the CPU memory of the corresponding node, as we have done in prior studies~\cite{Taylor2021}. The data required on each GPU is read once from a single NetCDF file containing the preprocessed data as described above. This approach facilitates the rapid loading of each batch of data to GPU memory and makes possible the highly scalable data-parallel training by preventing a filesystem IO bottleneck from occurring during training. This is particularly important when training the forecast CNN model as we construct a batch using a rolling window from the 12 past time steps and the future 2 time steps, so each sample in a batch consists of a total of 14 time steps. In order to further reduce memory usage for the  CNN model we define a data loader so we load from memory only the data that each batch requires at each time step.  We divide the training and test data sets equally by time onto each GPU. It is essential that each GPU has exactly the same number of time steps to avoid problems with load balancing and the timely communication of model parameters at the end of each epoch.

\section*{Results}

Having trained the CNN model we can then make forecasts (inferences) of the temporal evolution of the 2-D SST fields. The model inference step takes the preceding 12-months of SST data and makes predictions of the following two months. By using an autoregressive approach, where we feed model predictions back in as input to the model, we can make an unlimited number of predictions. For the results reported here we limit the predictions to a 24-month window. We make a series of 10 24-month predictions starting in July 2007 and commencing in each subsequent January and July, ending in January and July 2016. Each 24-month prediction period includes a 12-month overlap with the previous model predictions. Input data commences in January and July 2006, 12-months prior to the first predictions, and model predictions end in January and July 2019, 24-months after the start of the final prediction window. Model predictions span the data used both for model training and validation.

In this section, we first present results showing the predictions of the global scale SST fields. We also use the global scale SST fields to extract the SST values that correspond with well-known indices such as the ENSO index. As we are making predictions of the global scale fields we can extract any index of interest from our model predictions without the need to develop and train a new model.

\subsection*{Global scale 2-D SST Predictions}

\begin{figure}[h!]
\begin{center}
\includegraphics[width=17.5cm]{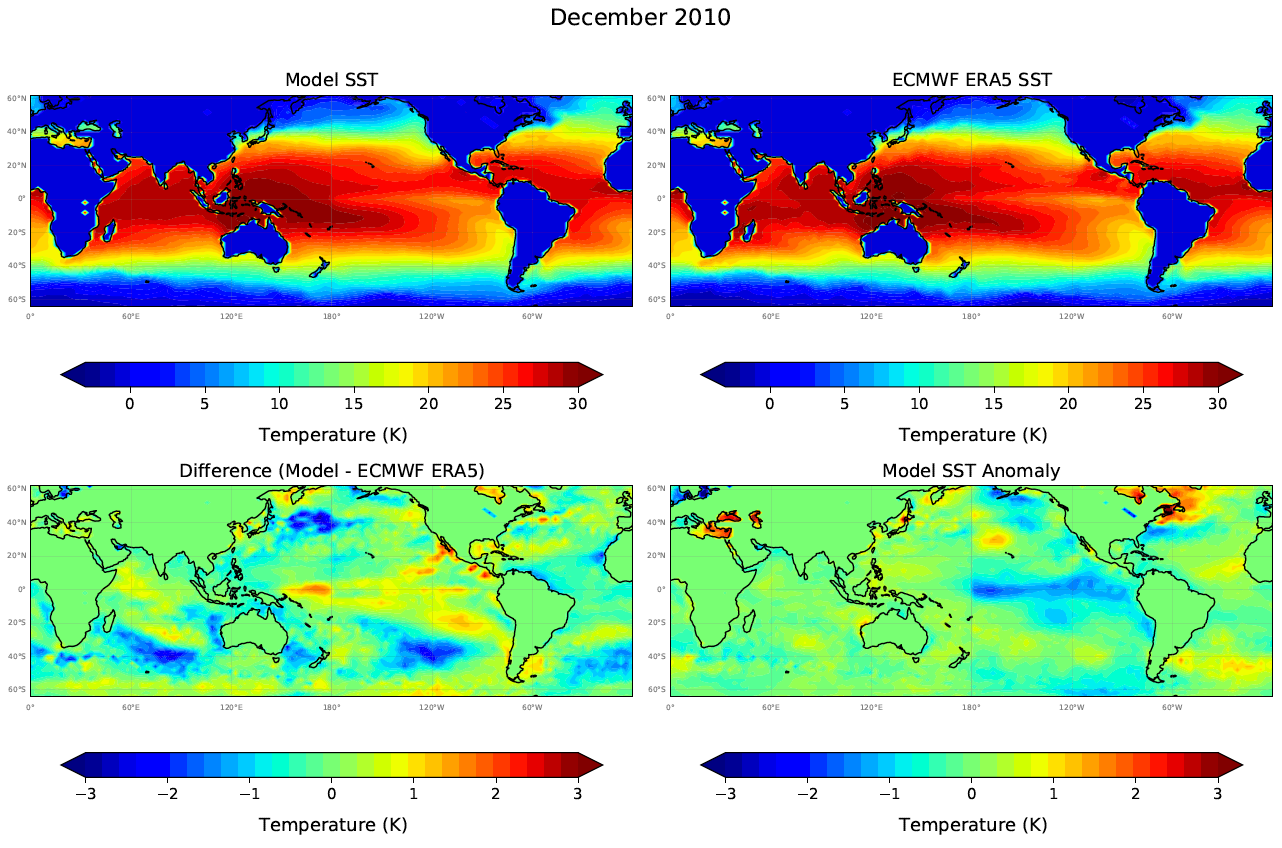}
\end{center}
\caption{Model predicted SST at +6 months (December 2010), the corresponding target ERA5 SST data set, the difference between the model predictions and the target, and the model predicted SST anomaly.  Note the different, much higher resolution scale used to plot model differences and SST anomalies.}\label{fig:4}
\end{figure}

We show an example set of SST model predictions at t=6,12,18, 24 months into the future starting in July 2010. Figure ~\ref{fig:4} presents the model predicted SST field at t=6, the corresponding target ECMWF ERA5 SST field, the difference between the model predicted and ERA5 SST, and the model predicted SST anomaly which is the difference between the model predicted SST values and an ERA5 climatology computed over the 30 year period 1981-2010. The model predicted SST values accurately capture the main features of the target ERA5 SST values with the majority of differences in SST values falling with the range $\pm 1 ^\circ$C. There does appear to be evidence that the model is systematically slightly warmer between $20 ^\circ$N and $20 ^\circ$S and slightly cooler at higher latitudes in Figure ~\ref{fig:4}. The model captures the SST cooling in the eastern equatorial Pacific during the developing 2010-11 La Ni\~{n}a, though with a slight warm bias, or underestimation of the cooling. Interestingly the model is also able to capture the warming SST off the northwest Australian coast at a 6-month lead, which is a precursor to the peak SST anomalies in February 2011 during extreme Ningaloo Ni\~{n}o event (Feng et al. 2013). Model predictions of warm SST anomalies in the eastern Mediterranean Sea and northwest Atlantic coast appear also being supported by observations (Figure ~\ref{fig:4}).

\begin{figure}[h!]
\begin{center}
\includegraphics[width=17.5cm]{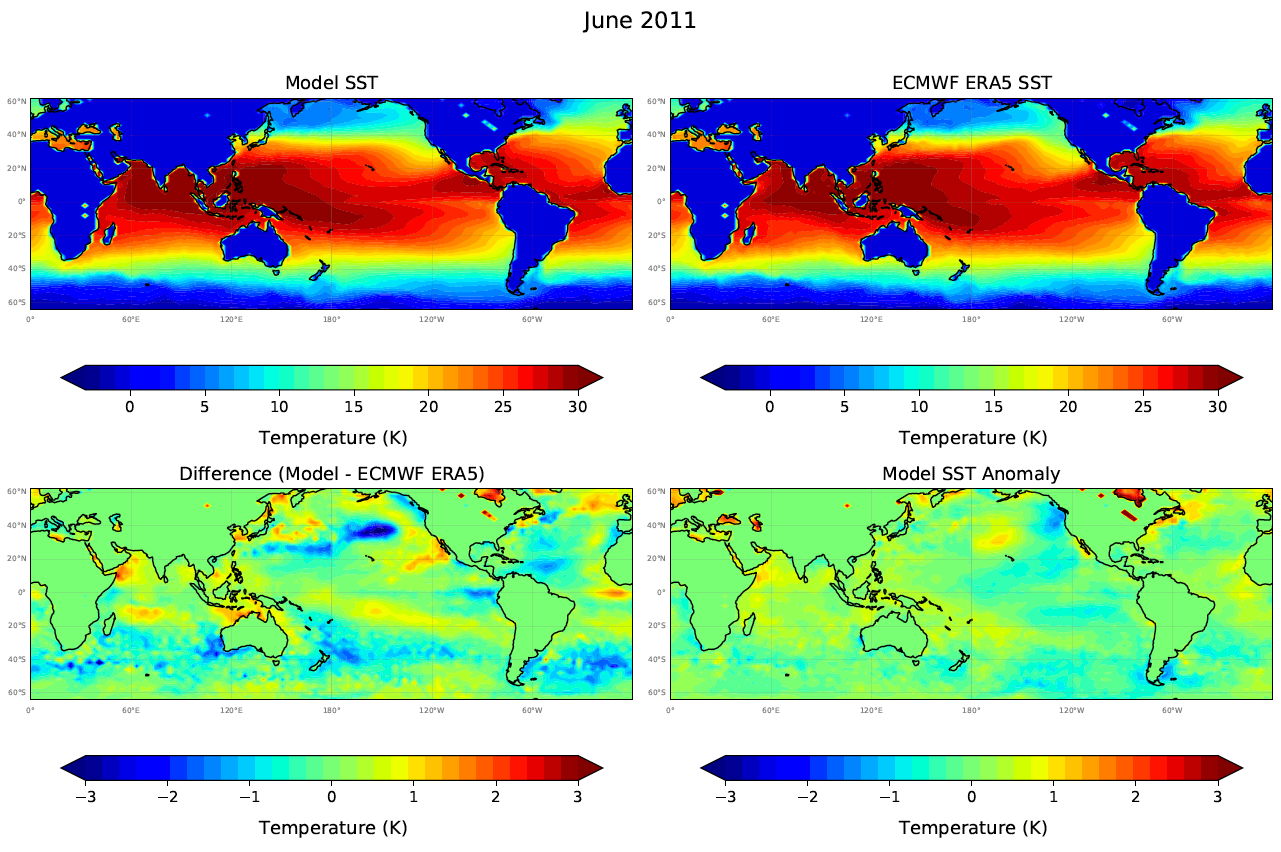}
\end{center}
\caption{Model predicted SST at +12 months (June 2011), the corresponding target ERA5 SST data set, the difference between the model predictions and the target, and the model predicted SST anomaly. Note the different, much higher resolution scale used to plot model differences and SST anomalies.}\label{fig:5}
\end{figure}

\begin{figure}[h!]
\begin{center}
\includegraphics[width=17.5cm]{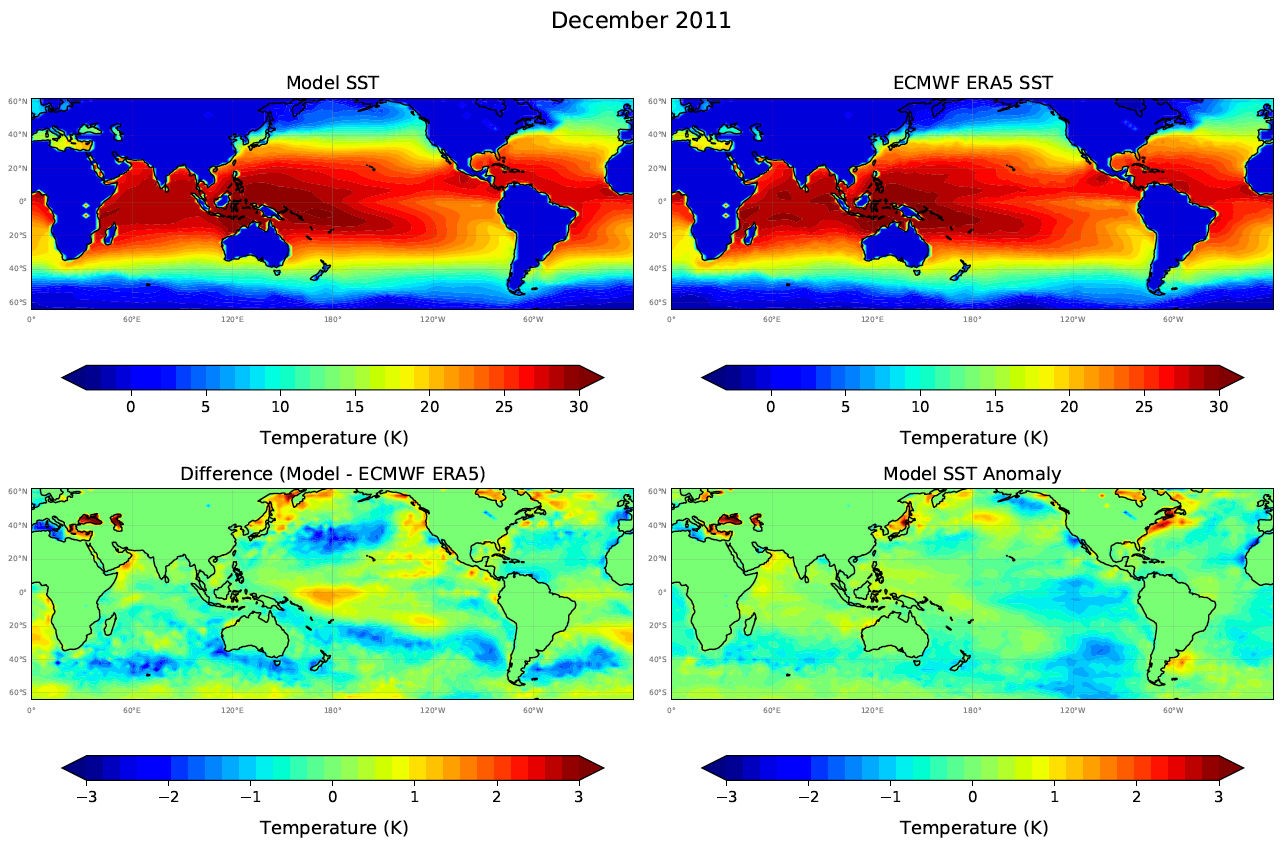}
\end{center}
\caption{Model predicted SST at +18 months (December 2011), the corresponding target ERA5 SST data set, the difference between the model predictions and the target, and the model predicted SST anomaly. Note the different, much higher resolution scale used to plot model differences and SST anomalies.}\label{fig:6}
\end{figure}

\begin{figure}[h!]
\begin{center}
\includegraphics[width=17.5cm]{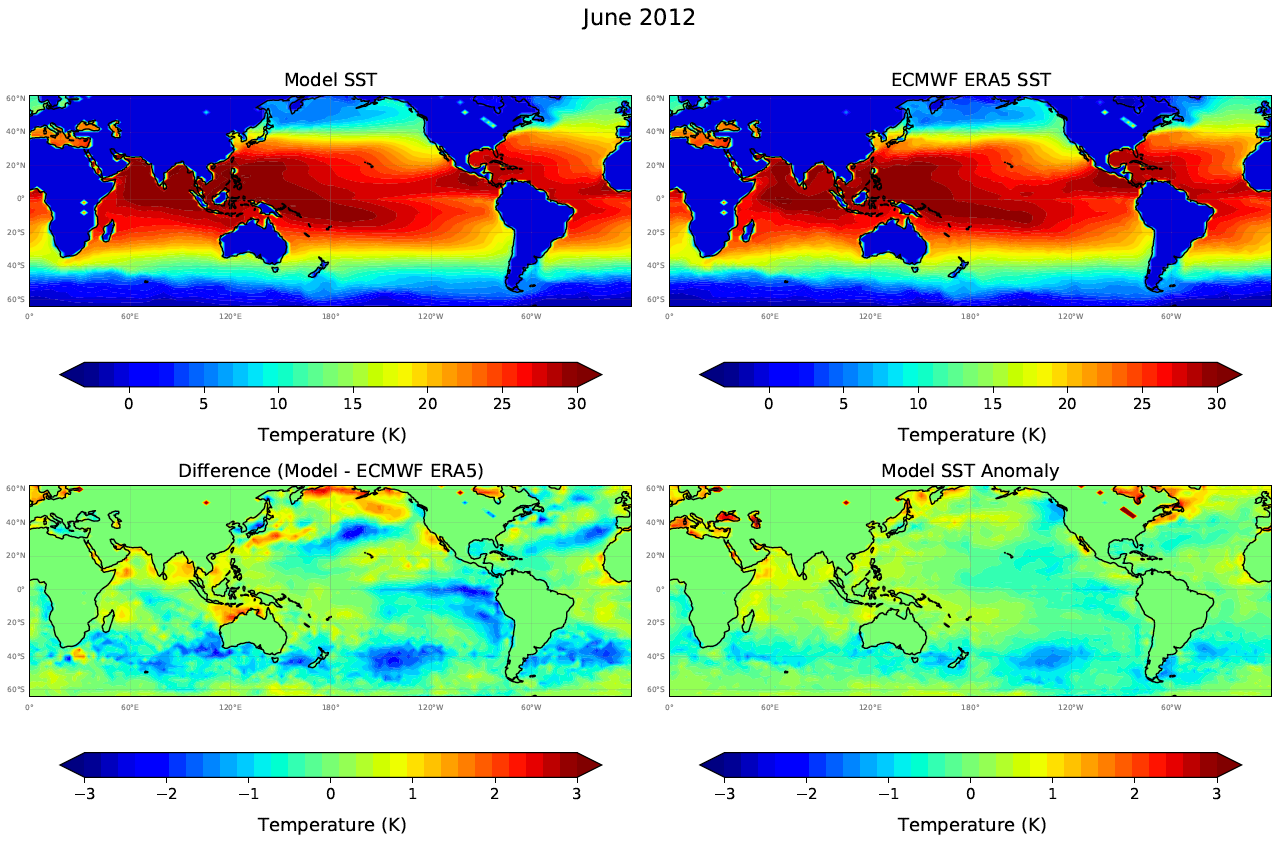}
\end{center}
\caption{Model predicted SST at +24 months (June 2012), the corresponding target ERA5 SST data set, the difference between the model predictions and the target, and the model predicted SST anomaly. Note the different, much higher resolution scale used to plot model differences and SST anomalies.}\label{fig:7}
\end{figure}

Figures ~\ref{fig:5} , ~\ref{fig:6}  and ~\ref{fig:7} compare CNN model predictions at t=12, 18 and 24 months into the future respectively. We can see that the model captures the temporal evolution of SST field well over the full 24 month prediction period. The differences between model predicted SST values grow slowly over the 24 month prediction period with the majority of differences in SST values, as shown in Figure~\ref{fig:7}, falling within the range $\pm 2 ^\circ$C with a small overall bias towards cooler temperatures. The RMSE value increases steadily over the 24 month prediction period from $0.43 ^\circ$C in July 2010 to $0.65 ^\circ$C in June 2012. At 18 month lead (Fig. 6), the CNN model predicts moderate cooling in the equatorial eastern Pacific, however, the model has a strong warm bias near the dateline. Thus there appears to be some skill to predict the 2011-12 moderate La Ni\~{n}a at the 18-month lead, though the cooling extends less westward compared with observation due to some warm bias. The anomaly features predicted at mid- to high-latitude at 18-month lead are mostly due to model errors (Figure ~\ref{fig:6}). Nevertheless, the CNN captures the underlying seasonal variations of the global SST quite accurately.

\begin{figure}[h!]
\begin{center}
\includegraphics[width=17.5cm]{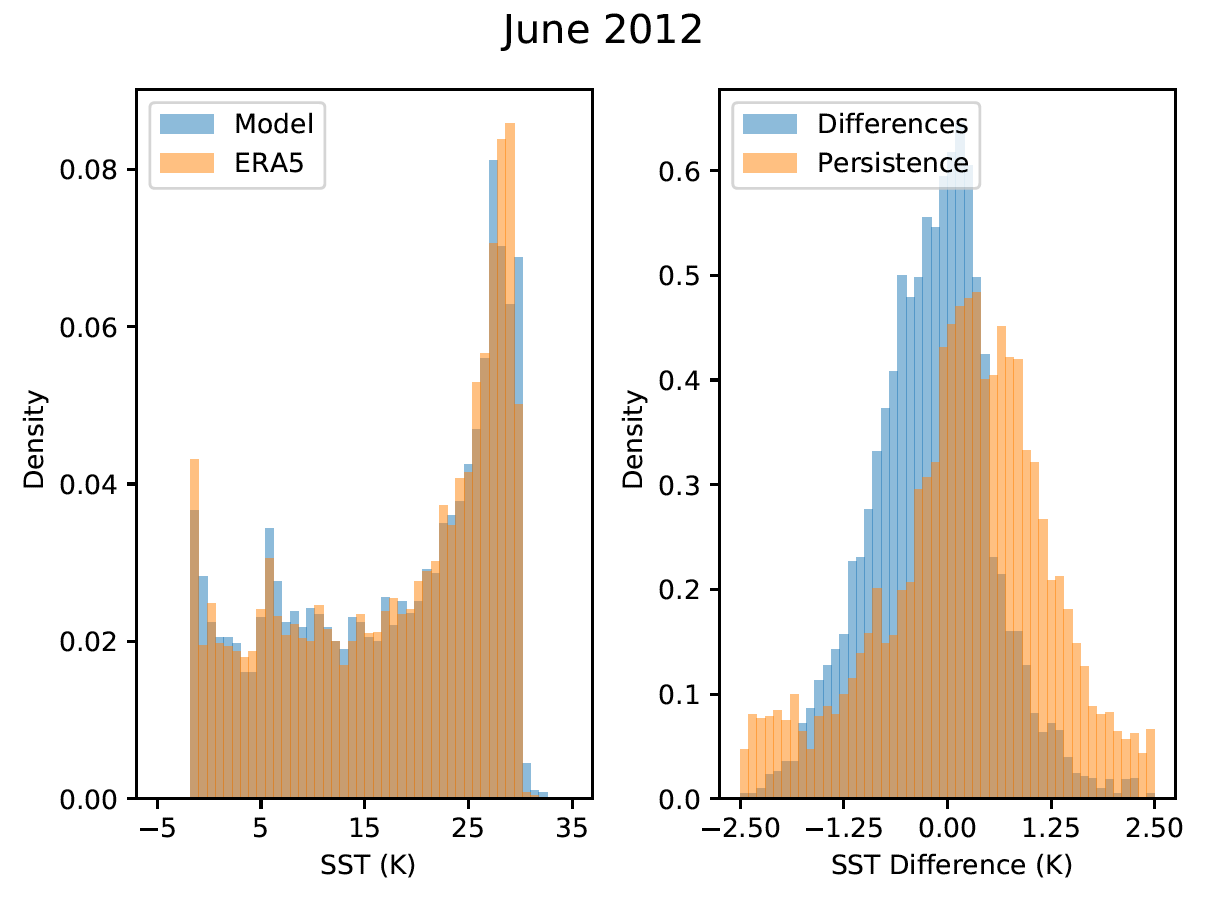}
\end{center}
\caption{Histograms of the model predicted SST values in comparison with the ERA5 SST values for June 2012 at the end of the 24 month prediction period.  The second panel shows a histogram of the differences between the model and the ERA5 SST values in comparison with errors produced by assuming persistence.}\label{fig:8}
\end{figure}

Figure ~\ref{fig:8} presents histograms of the CNN model predicted SST values in comparison with the ERA5 SST values for June 2012, corresponding to the results presented in Figure~\ref{fig:7}, at the end of the 24 month prediction period. Figure ~\ref{fig:8} clearly demonstrates that the model is able to accurately maintain the correct distribution of SST values with no smearing of the distribution even at the end of the 24 month prediction window. This can be attributed to the use of the Conv2DLSTM layers in the CNN model which correctly capture both the spatial and temporal evolution of the 2D SST field. The second panel in Figure ~\ref{fig:8} shows a histogram of the differences between the model and the ERA5 SST values in comparison with errors produced by assuming persistence from June 2010. The histogram of the model differences is centred close to $0 ^\circ$C with the majority of errors falling within the range $\pm 2 ^\circ$C, as previously shown in Figure~\ref{fig:7}. 

\begin{figure}[h!]
\begin{center}
\includegraphics[width=17.5cm]{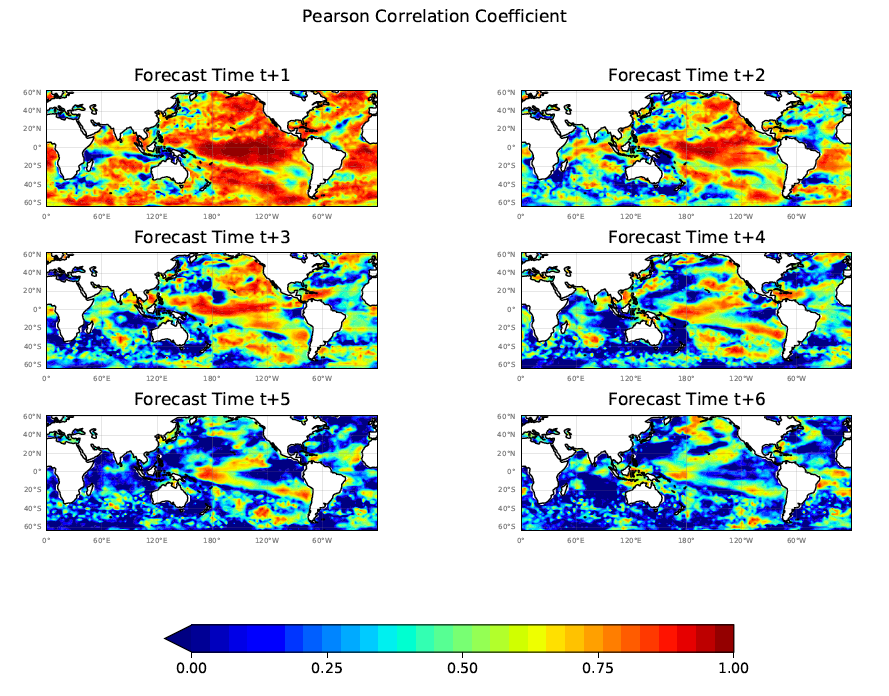}
\end{center}
\caption{The Pearson correlation coefficient (R) calculated using the 10 24-month forecasts commencing in July 2006 for months t+1 to t+6 showing large regions of significant correlation between the model predicted and ERA5 SST values out to t+6.} \label{fig:9}
\end{figure}

\begin{figure}[h!]
\begin{center}
\includegraphics[width=17.5cm]{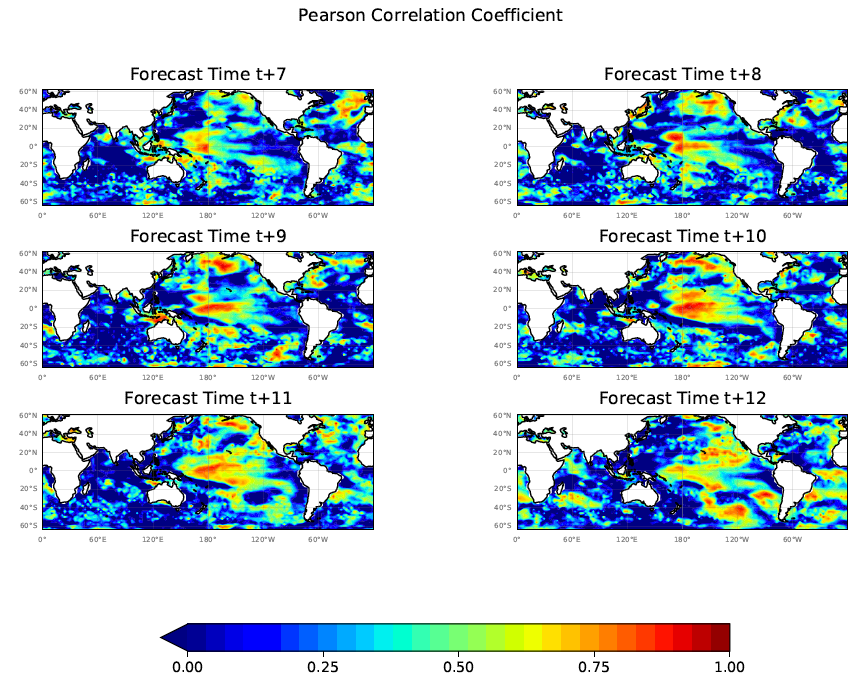}
\end{center}
\caption{The Pearson correlation coefficient (R) calculated using the 10 24-month forecasts commencing in July 2006 for months t+7 to t+12 showing the presence of significant correlation between the model predicted and ERA5 SST values out to t+12.} \label{fig:10}
\end{figure}

Figure ~\ref{fig:9} shows the estimates of the Pearson correlation index for our predicted SST against the ERA5 SST data based on the 10 24-month predictions starting in July 2006 for months t+1 to t+6. Figure ~\ref{fig:10} shows a corresponding plot to Figure ~\ref{fig:9} except for the months t+7 to t+12. These figures illustrate that the CNN model is able to maintain a significant correlation with the target ERA5 SST data for predictions out to t+12 months. Not shown are plots of the Pearson correlation index for t+13 to t+24 which continue to show regions of significant correlation. In general, long-lead high prediction skills are mostly located in the tropical, northeast, and south Pacific. There are also high prediction skills for the high latitude North Atlantic. There are good skills for the Indian Ocean Dipole regions up to 3-month lead, and the skills decay rapidly, likely due to a winter prediction barrier of the IOD (e.g. Luo et al. 2007).

\subsection*{Long-lead Predictions of the El Ni\~{n}o 3.4 and El Ni\~{n}o 4 indices}

Using the CNN predicted SST values we can calculate the El Ni\~{n}o 3.4 index computed over the region $5 ^\circ$S-$5 ^\circ$N, $170 ^\circ$W-$120 ^\circ$W.  We compare the model predictions with the El Ni\~{n}o 3.4 index derived from the ERA5 SST data.  The ERA5 El Ni\~{n}o 3.4 index is defined as the difference between the monthly mean ERA5 SST values and an ERA5 climatology computed over the 30 year period 1981-2010. 

\begin{figure}[h!]
\begin{center}
\includegraphics[width=12.5cm, angle=-90]{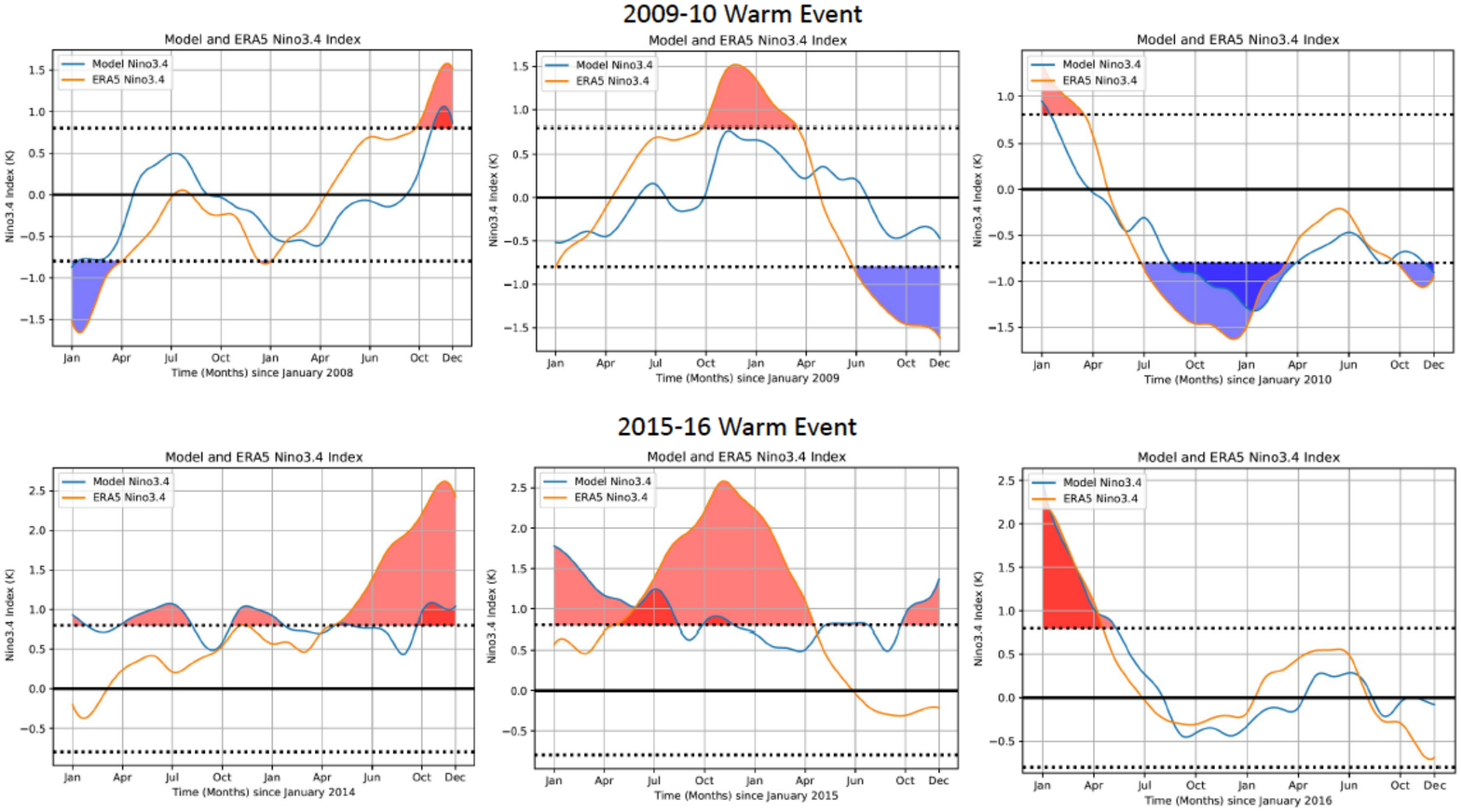}
\end{center}
\caption{Model predictions of the  El Ni\~{n}o 3.4 index for the 2009-10 and 2015-16 warm periods. For each warm period we show a sequence of three 24 month predictions starting 1 year apart that span the warm period. In each graph we compare the model results with the El Ni\~{n}o 3.4 index, defined as the monthly average SST computed over the region $5 ^\circ$N-$5 ^\circ$S, $120 ^\circ$W-$170 ^\circ$W   , calculated from the monthly mean ERA5 SST data. The dotted lines define warm ($>0.8 ^\circ$C) and cold ($<-0.8 ^\circ$C) periods. Model predictions and ERA5 estimates of the  El Ni\~{n}o 3.4 index $>0.8 ^\circ$C have been shaded red, and values $<-0.8 ^\circ$C have been shaded blue, for emphasis.}\label{fig:11}
\end{figure}

We have performed a series of ten 24 month predictions of the El Ni\~{n}o 3.4 index starting each January, commencing in January 2008, to assess the model's ability to overcome the spring barrier in predicting El Nino. The model predictions started from July have generally improved skill. Figure ~\ref{fig:11} shows the model predicted the  El Ni\~{n}o 3.4 index for the 2009-10 and 2015-16 warm periods. For each warm period we show a sequence of three 24 month predictions starting 1 year apart that span the warm period. The first panel in the sequence shows the 24 month period leading up to the warm event, the second panel spans the warm event and the panel plot shows the 24 months exiting the warm event. 

For the 2009-10 warm event, the top three panels in Figure ~\ref{fig:11}, the predicted El Ni\~{n}o 3.4 index closely follows the ERA5 El Ni\~{n}o 3.4 index over the 24 months leading to the warm event, with the warm event correctly predicted 24 months in advance. The predictions starting in January 2008 spanning the 2009-10 warm event, starting a relatively cold state, capture the strong warming through 2009 but do underestimate the peak in the ERA5 El Ni\~{n}o 3.4 index by $0.7 ^\circ$C. Similarly, the predictions starting in January 2009 capture a warming peak during the 2009-10 boreal winter, as well as the transition toward a cooler state afterwards. The 24 month prediction of the exit from the warm period, starting in January 2010, very closely follows the ERA5 El Ni\~{n}o 3.4 index and correctly predicts the strong La Ni\~{n}a cooling event in 2010-11.

For the 2015-16 warm event, the lower three panels in Figure ~\ref{fig:11}, the predicted El Ni\~{n}o 3.4 index consistently predicts warmer conditions throughout the 24 month prediction period, to some extent due to the persistence from the initial condition in the case of January 2014, but does not capture the strength of the 2015-16 warm event. The prediction starting from January 2015 again shows warm conditions throughout the 24 month forecast though failing to capture the full intensity of the event commencing in April 2015 and the transition toward a cooler state in 2016. The model appears not be able to overcome the spring prediction barrier in this occasion, even starting from a relatively warm state in January 2015. As we observed for the 2009-10 warm event, the 24 month prediction of the exit from the 2015-16 warm event very closely follows the ERA5 El Ni\~{n}o 3.4 index. The model correctly predicts the strong decline in ERA5 El Ni\~{n}o 3.4 index from its peak in January 2016 and ongoing normal conditions throughout 2017.

\begin{figure}[h!]
\begin{center}
\includegraphics[width=12.5cm, angle=-90]{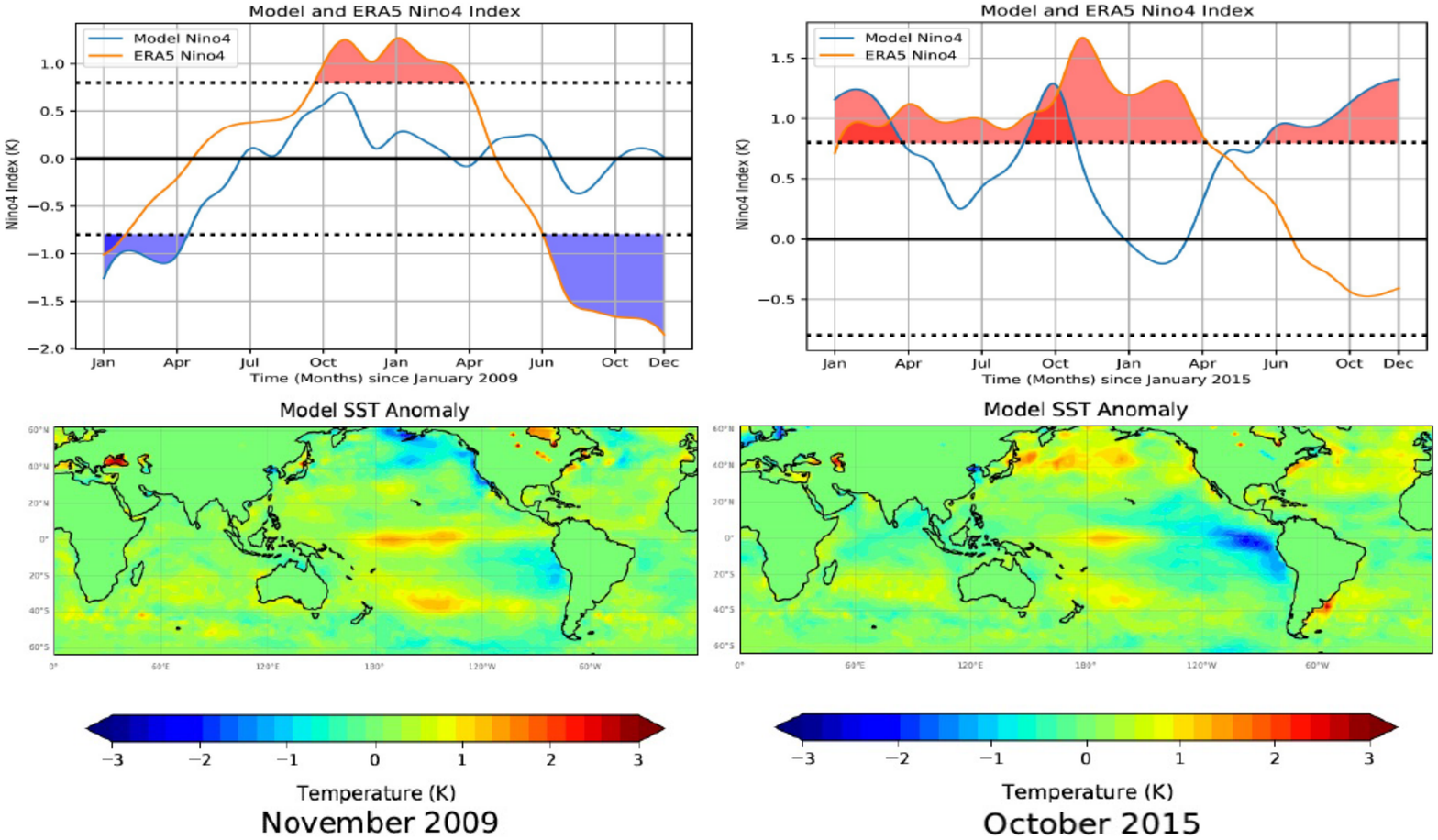}
\end{center}
\caption{Model predictions of the  El Ni\~{n}o 4 index for the 2009-10 and 2015-16 warm periods are presented in the upper panels. In each graph we compare the model results with the El Ni\~{n}o 4 index, defined as the monthly average SST computed over the region $5 ^\circ$N-$5 ^\circ$S, $160 ^\circ$E-$150 ^\circ$W, calculated from the monthly mean ERA5 SST data. The dotted lines define warm ($>0.8 ^\circ$C) and cold ($<-0.8 ^\circ$C) periods. Model predictions and ERA5 estimates of the  El Ni\~{n}o 4 index $>0.8 ^\circ$C have been shaded red, and values $<-0.8 ^\circ$C have been shaded blue, for emphasis. The lower panels present the spatial pattern of the model predicted SST anomalies in November 2009 and October 2015 taken from the corresponding 24 month predictions shown in the upper panels.}\label{fig:12}
\end{figure}

As the 2009-10 El Ni\~{n}o is generally regarded as a central Pacific El Ni\~{n}o ~\cite{Timmermann2018}, we also assess the prediction of Ni\~{n}o 4 SST variability, an index for the central Pacific warming. The Ni\~{n}o 4 index is computed over the region $5 ^\circ$S-$5 ^\circ$N, $160 ^\circ$E-$150 ^\circ$W. Figure ~\ref{fig:12}, upper panels, show the ERA5 Ni\~{n}o 4 index over the 24 months covering the 2009-10 warm event and the 2015-16 warm event. The lower panels in Figure ~\ref{fig:12} presents the spatial pattern of the model predicted SST anomalies in November 2009 and October 2015 taken from the corresponding 24 month Ni\~{n}o 4 index predictions in the panels above. It is shown that the model can predict the Ni\~{n}o 4 index variability well, though with an early peak in the predicted SST anomalies in October 2009 (Figure ~\ref{fig:12}), the warming pattern is similar to a central El Ni\~{n}o event (Figure ~\ref{fig:12}). The longer lead prediction starting from January 2008 show similar results (not shown). 

\begin{figure}[h!]
\begin{center}
\includegraphics[width=12.5cm, angle=-90]{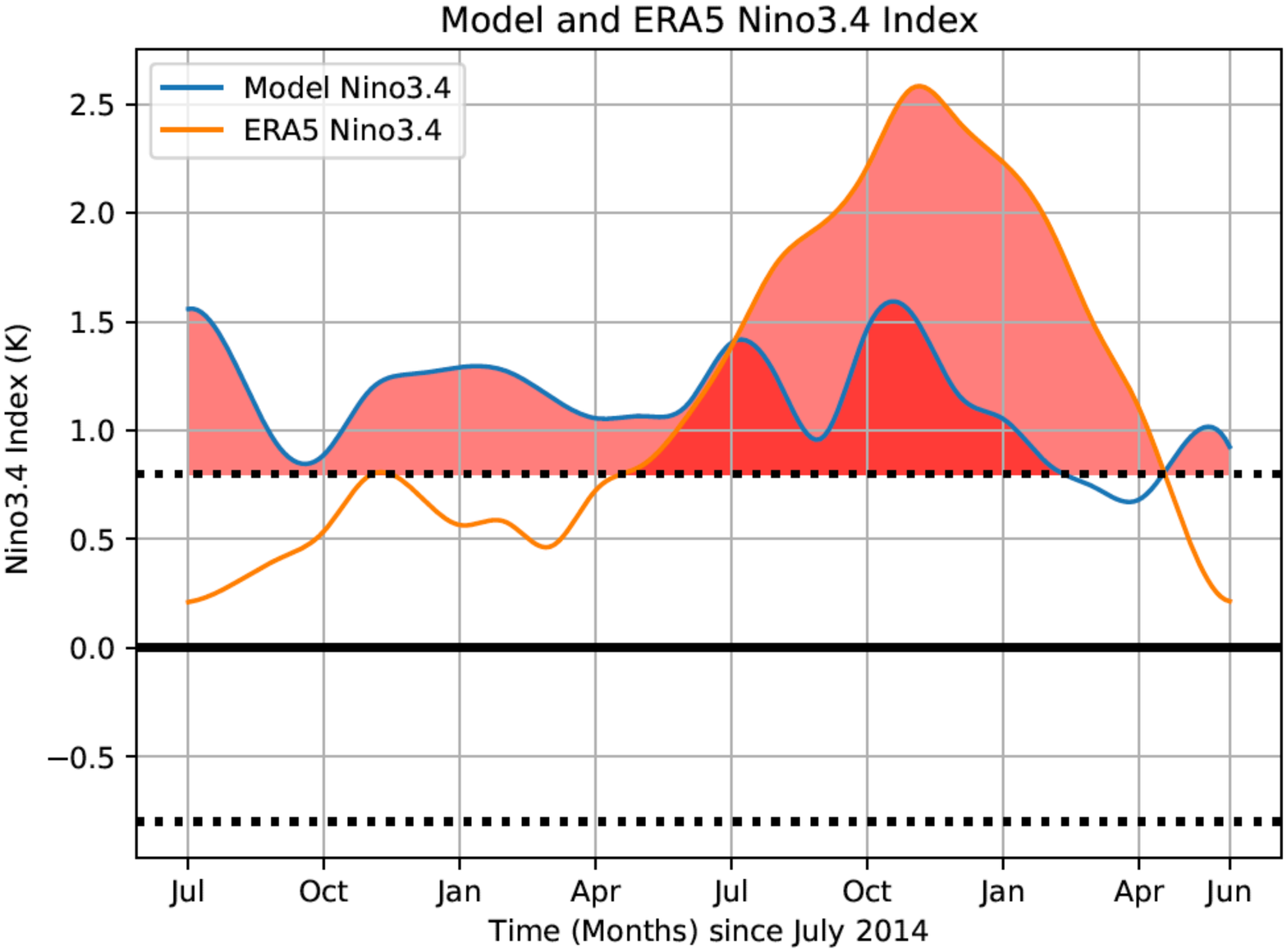}
\end{center}
\caption{Model predictions of the  El Ni\~{n}o 3.4 index for a 24 month prediction starting in July 2014 and ending in June 2016. We compare the model results with the El Ni\~{n}o 3.4 index, defined as the monthly average SST computed over the region $5 ^\circ$N-$5 ^\circ$S, $120 ^\circ$W-$170 ^\circ$W   , calculated from the monthly mean ERA5 SST data. The dotted lines define warm ($>0.8 ^\circ$C) and cold ($<-0.8 ^\circ$C) periods. Model predictions and ERA5 estimates of the  El Ni\~{n}o 3.4 index $>0.8 ^\circ$C have been shaded red, and values $<-0.8 ^\circ$C have been shaded blue, for emphasis.}\label{fig:13}
\end{figure}

The model prediction correctly captures an early central Pacific warming in Oct’15, but it quickly decays. The model fails to predict the sustained warming in the eastern equatorial Pacific, instead, it predicts a transition to La Nina cooling (Figure ~\ref{fig:12}. This is likely due to that it is rare in the training data that has two consecutive El Ni\~{n}os, with the second one being stronger. Historically, most El Ni\~{n}o events are followed by La Ninas. Figure ~\ref{fig:13} shows model predictions of the  El Ni\~{n}o 3.4 index for a 24 month prediction starting in July 2014 (rather than in January) and ending in June 2016 which spans the entire two year warm period. Figure ~\ref{fig:13}  illustrates that the model is able to capture this unusual event with two consecutive El Ni\~{n}os with model predictions tracking the ERA5 El Ni\~{n}o 3.4 index within $~1 ^\circ$C, throughout the full two year period, though it still does not match the full intensity of the 2015-16 El Ni\~{n}o. As demonstrated in the correlation maps, we note that most of the ENSO events can be well captured by the Unet-LSTM model.

\subsection*{Long-lead Predictions of the 'Blob' index}

\begin{figure}[h!]
\begin{center}
\includegraphics[width=17.5cm]{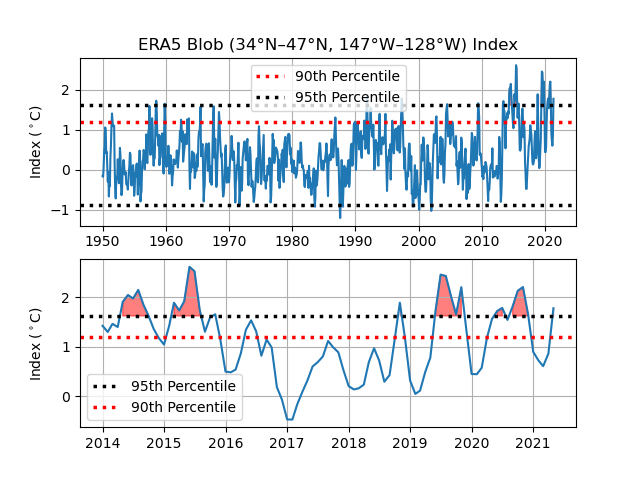}
\end{center}
\caption{The Blob index defined as the difference between the monthly average SST climatology (1980-2010) and the monthly average SST computed over the region $34 ^\circ$N-$47 ^\circ$N, $147 ^\circ$W-$128 ^\circ$W. We compute the Blob index from the monthly mean ERA5 SST data for the period January 1950-May 2021. We include the two-sided 95\% confidence interval and the one-sided 90th percentile. The lower panel is the same Blob index with a focus on the period January 2014-May 2021 with the significant ocean warming events in 2014, 2015, 2019 and 2020 highlighted.}\label{fig:14}
\end{figure}

We have computed the Blob index using the ERA5 SST data for the period January 1950-May 2021. The Blob index is defined as the difference between the monthly average SST climatology (1980-2010) and the monthly average SST computed over the region $34 ^\circ$N-$47 ^\circ$N, $147 ^\circ$W-$128 ^\circ$W. Figure ~\ref{fig:14} presents the blob index over this time period, along with a two-sided 95\% confidence interval and a one-sided 90\% confidence interval. In the lower panel of Figure ~\ref{fig:14} we focus on the period January 2014-May 2021.  During this time period we see multiple exceptional ocean warming events in the Blob region where ocean temperatures exceed the 95\% confidence interval by up to $1 ^\circ$C. It is also worth noting during the January 2014-May 2021 period that the Blob index only dropped below $0 ^\circ$C during the boreal winter of 2016-17.

\begin{figure}[h!]
\begin{center}
\includegraphics[width=12.5cm, angle=-90]{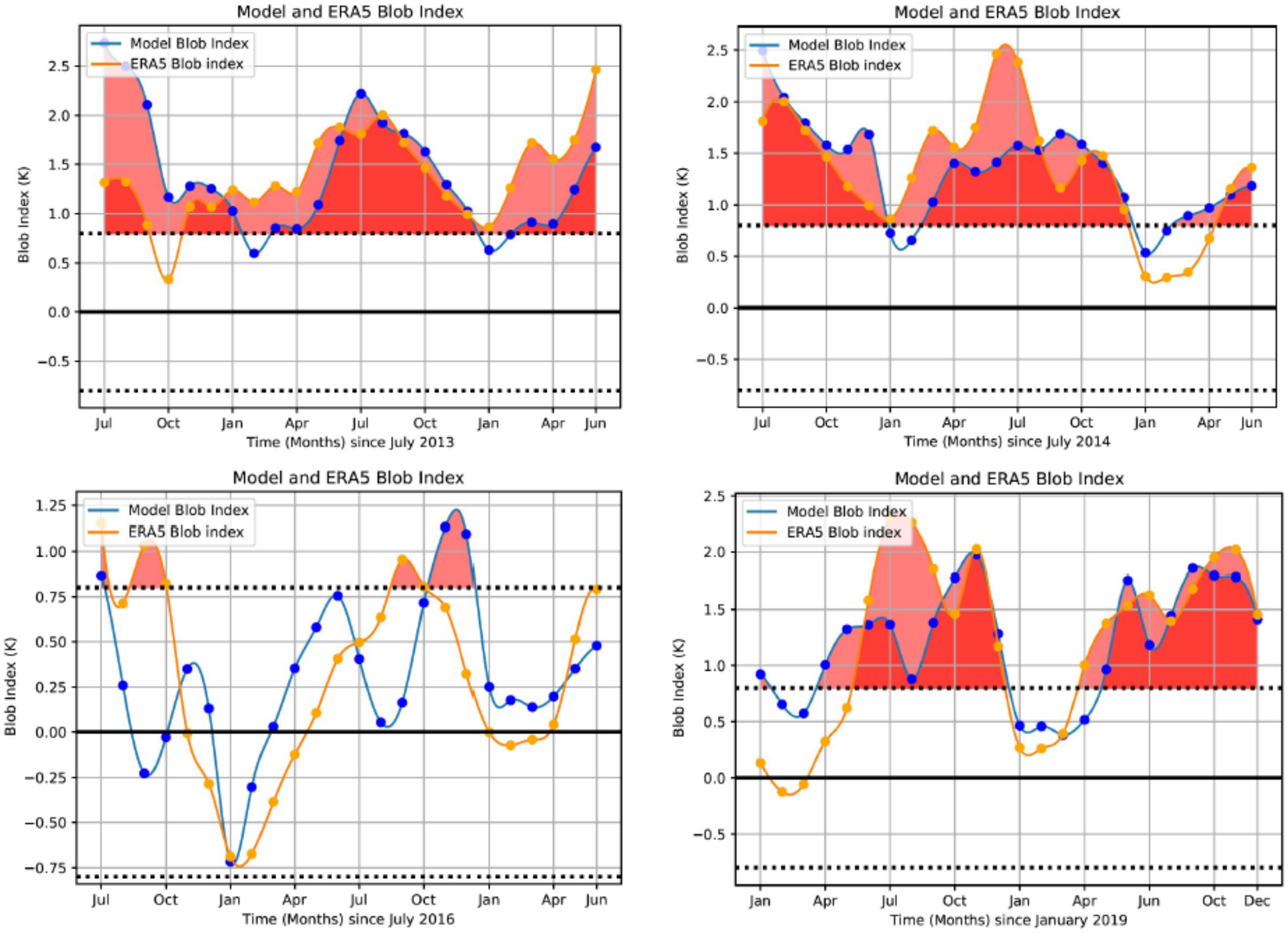}
\end{center}
\caption{Model predictions of the Blob index over a 24 month period commencing in July 2013, 2014, 2015 and January 2019 compared with the Blob index, defined as the monthly average SST computed over the region $34 ^\circ$N-$47 ^\circ$N, $147 ^\circ$W-$128 ^\circ$W, calculated from the monthly mean ERA5 SST data. We include the two-sided 95\% confidence interval. We focus on the period January 2014-May 2021 with the significant ocean warming events in 2014 and 2015 captured in the top two panels. The winter cooling in 2016-17 and the warming events in 2019 and 2020 are presented in the lower panels.}\label{fig:15}
\end{figure}

In order to demonstrate the ability of the CNN model to predict the Blob index Figure ~\ref{fig:15} presents multiple 24 month predictions of the Blob index focusing on the period 2014-21. The Unet-LSTM model Blob index is computed from the corresponding model predicted SST values e.g. as presented in Figures ~\ref{fig:4}-~\ref{fig:7} and monthly average SST climatology (1980-2010). We include 24 month predictions that capture the warm events in 2014 and 2015, the cooling during the winter of 2016-17 and the two warm events in 2019 and 2020. Figure ~\ref{fig:15} illustrates the ability of the model to accurately predict the evolution of the Blob index during both warming and cooling events over the full 24 month prediction period.

\section*{Summary and Discussion}

In this study, we use monthly reanalysis global surface temperature (SST and 2-m air temperature) data to train a Unet-LSTM  data-driven model and demonstrate its ability to predict SST variability at various lead times. We used a 12-month window to train the Unet-LSTM model, with the seasonal cycles retained in the training data, which effectively captured the seasonal SST variation in the global ocean. For the SST anomaly prediction, there are high long-lead skills in the equatorial Pacific and northeast Pacific. In the following, we discuss a few aspects of the model predictions and outline our plans for future work. 

\subsection*{Comparison with other ML ENSO prediction architecture}

~\cite{Ham2019} developed a CNN deep learning model, trained with coupled climate model outputs, to achieve an 18 month-lead prediction skills for Nino3.4. Their model was initialized with both monthly SST and upper ocean heat content anomalies, claiming that the upper ocean heat content memory actually helped the model to achieve the long-lead model skills. Most of the recent development in deep learning ENSO forecasting model are based on this framework (~\cite{Ham2021} among others). The Unet-LSTM based CNN model trained and initialized with global surface temperature fields can achieve similar prediction skills, not only for Nino3.4, but also in the northeast Pacific, which is demonstrated in the prediction assessment of the recent Blob marine heatwave events. 

We have used a 12-month window to train the Unet-LSTM model, while most other CNN models used 3-month temporal window. It appears that the upper ocean memory may also reside in the surface temperature records. Surface SST and land surface temperatures drive global surface wind anomalies and then subsequently drive the planetary ocean waves to store the upper ocean heat content anomalies. This is a problem for further exploration. The model achieves long-lead prediction mostly in the Pacific, which suggests that the precursors of the long-lead prediction likely reside in the Pacific.

By using a 12-month time window to train the Unet-LSTM model, we can use the full temperature field, instead of only using the anomaly field. In this way, both the annual cycle of temperature variations and the interannual anomalies are considered simultaneously. This may have two benefits: one is to be able to train the model to assimilate the dynamics of the seasonally phase-locked variability; the other is that the model can carry the memory over the past years, so that it is not necessary to remove the steady warming trend at the surface ocean from the reanalysis (observation) data prior to the model training.

Note that there is an attempt to capture the seasonal cycle by introducing additional labels in a CNN model ~\cite{Ham2021}, however, it may only be achievable for a single index prediction. In the results, we mostly show the prediction starting from January to demonstrate the model ability to overcome the spring prediction barrier.  We do need to assess the stability of the model prediction starting from different months, especially when there is known prediction barriers for various climate indices (e.g. ~\cite{Timmermann2018}). While the Unet-LSTM model is able to predict most ENSO events well, it is noted that the model prediction starting in January fails to predict the 2015-16 peak during the rare occurrence of two consecutive El Ninos, likely due to lack of similar cases in the training data based on existing observations. ENSO diversity may still pose a challenge for the data-driven machine learning models. Nevertheless, the current version of Unet-LSTM shows great promise in leading the way to more sophisticated 2-dimensional SST predictions for the global ocean.

\subsection*{Future work}

The success of the Unet-LSTM model at capturing key features of the global scale temperature field clearly demonstrates that data driven approaches to modelling the spatio-temporal evolution of complex physical systems are a promising avenue for further research. As a next step we plan to increase the model resolution to the full resolution of the ERA5 data set, currently $0.25^\circ$, which will allow us to investigate the impact of model resolution on SST predictions and to study the variations in SST at the regional scale in more detail. We will also investigate better quantifying the uncertainty associated with model forecasts using an ensemble forecasting approach and the prediction of other key ocean indices, such as the Indian Ocean Dipole (IOD), using both the existing model and at higher model resolutions.

Given the results presented here and the successful application of the Unet-LSTM model in previous studies~\cite{Taylor2021} we have increased confidence that the Unet-LSTM model can be applied to the general problem of the spatial and temporal evolution of other 2D geophysical fields. As of TensorFlow 2.6, a ConvLSTM3D layer is now available. By replacing the ConvLSTM2D layer with the ConvLSTM3D layer, the Unet-LSTM  can be used to predict the spatio-temporal evolution of 3D fields. The ability to model 3D fields will allow us to investigate improving SST predictions by incorporating additional input variables, such as surface wind fields, into the Unet-LSTM model. The primary barrier to working with large 3D fields is the availability of GPU memory which on current devices is limited to 16-32GB. Next generation GPU devices will have significantly larger memory, which combined with model parallelism, will allow much larger more complex models to be developed.

Training the Unet-LSTM model only with the reanalysed temperature field (which incorporates existing observations) over the seven decades appears to have constrained the model to capture the ENSO dynamics and its teleconnection in the Indian Ocean and mid-latitude oceans (e.g. the Blob region). On the other hand, the CNN ML models are trained using coupled atmosphere-ocean models, which have inherent biases in the coupled models, as well as unrealistic ENSO simulations in some of the coupled models, such as the ENSO frequencies. Transfer learning, which has been proposed in some studies~\cite{Ham2019}, may not be enough to correct these model biases~\cite{Timmermann2018}.  A knowledge based strategy is needed to combine the coupled model results with observations to provide a well sampled dataset, for the machine learning models to capture the diverse SST variability in the global and regional oceans, in order to better predict rare climate events.

A new and rapidly evolving area of research is physics-informed machine learning ~\cite{Karniadakis2021} that combines machine learning with physical constraints, derived for example, from ordinary differential equations (ODEs) and partial differential equations (PDEs) that describe the system under study. The implementation of physics-informed machine learning is mesh-less which allows model regression to take place using an available set of imperfect observations that define the initial and boundary conditions without the need to interpolate the data to an appropriate grid. Physics-informed machine learning could yield new more flexible, potentially transformative, approaches to ocean modelling however much work needs to be done to realise this goal.

\section*{Conflict of Interest Statement}

The authors, JT and MF, declare that the research was conducted in the absence of any commercial or financial relationships that could be construed as a potential conflict of interest.

\section*{Author Contributions}

The authors, JT and MF, have contributed equally to this work.

\section*{Acknowledgments}
This research project was undertaken with the assistance of the resources and services from the National Computational Infrastructure (NCI), which is supported by the Australian Government.

\section*{Data Availability Statement}
The datasets generated for this study can be found in the CSIRO Data Access Portal https://data.csiro.au.

\bibliographystyle{chicago}
\bibliography{arXiv_JAT_MF.bib}


\pagebreak

\section*{Figures}

\end{document}